\documentclass[9pt,conference]{IEEEtran}

\usepackage[utf8]{inputenc} 
\usepackage[T1]{fontenc}    
\usepackage{url}            
\usepackage{booktabs}       
\usepackage{amsfonts}       
\usepackage{nicefrac}       
\usepackage{microtype}      

\usepackage{pgfplots}
\usetikzlibrary{positioning,shapes,arrows,patterns,fit,backgrounds,calc,svg.path,3d,decorations.text,shapes.arrows,spy}
\usepackage{todonotes}
\usepackage{graphicx}
\usepackage{subfigure}
\usepackage{amsmath,amssymb}
\usepackage{array}
\usepackage{makecell}
\usepackage{caption}
\usepackage{multirow}
\usepackage[hang,flushmargin]{footmisc}

\DeclareRobustCommand*{\IEEEauthorrefmark}[1]{%
  \raisebox{0pt}[0pt][0pt]{\textsuperscript{\footnotesize #1}}%
}

\pgfdeclarelayer{background}
\pgfdeclarelayer{front}
\pgfsetlayers{background,main,front}

\makeatletter
\pgfkeys{%
	/tikz/on layer/.code={
		\pgfonlayer{#1}\begingroup
		\aftergroup\endpgfonlayer
		\aftergroup\endgroup
	},
	/tikz/node on layer/.code={
		\gdef\node@@on@layer{%
			\setbox\tikz@tempbox=\hbox\bgroup\pgfonlayer{#1}\unhbox\tikz@tempbox\endpgfonlayer\egroup}
		\aftergroup\node@on@layer
	},
	/tikz/end node on layer/.code={
		\endpgfonlayer\endgroup\endgroup
	}
}

\def\node@on@layer{\aftergroup\node@@on@layer}

\pgfplotsset{ every non boxed x axis/.append style={x axis line style=-},
	every non boxed y axis/.append style={y axis line style=-}}

\def\BibTeX{{\rm B\kern-.05em{\sc i\kern-.025em b}\kern-.08em
		T\kern-.1667em\lower.7ex\hbox{E}\kern-.125emX}}

\begin{document}
	
\title{Complex-valued Convolutional Neural Networks for Enhanced Radar Signal Denoising and Interference Mitigation}

\author{
	\IEEEauthorblockN{
	Alexander Fuchs\IEEEauthorrefmark{1}\textsuperscript{\textsection},
	Johanna Rock\IEEEauthorrefmark{1}\textsuperscript{\textsection},
	Mate Toth\IEEEauthorrefmark{1}\IEEEauthorrefmark{2},
	Paul Meissner\IEEEauthorrefmark{2}, 
	Franz Pernkopf\IEEEauthorrefmark{1}}
	\IEEEauthorblockA{\IEEEauthorrefmark{1}Graz University of Technology, Austria, \IEEEauthorrefmark{2}Infineon Technologies Austria AG, Graz}
	\IEEEauthorblockA{Email: fuchs@tugraz.at, johanna.rock@tugraz.at}
}


\maketitle

\begingroup
\renewcommand\thefootnote{\textsection}
\footnotetext{Both authors contributed equally.\\
This work was supported by the Austrian Research Promotion Agency (FFG) under the project SAHaRA (17774193) and by NVIDIA by providing GPUs.}
\endgroup

\begin{abstract}
Autonomous driving highly depends on capable sensors to perceive the environment and to deliver reliable information to the vehicles' control systems. To increase its robustness, a diversified set of sensors is used, including radar sensors. Radar is a vital contribution of sensory information, providing high resolution range as well as velocity measurements. The increased use of radar sensors in road traffic introduces new challenges. As the so far unregulated frequency band becomes increasingly crowded, radar sensors suffer from mutual interference between multiple radar sensors. This interference must be mitigated in order to ensure a high and consistent detection sensitivity.
In this paper, we propose the use of Complex-Valued Convolutional Neural Networks (CVCNNs) to address the issue of mutual interference between radar sensors. We extend previously developed methods to the complex domain in order to process radar data according to its physical characteristics. This not only increases data efficiency, but also improves the conservation of phase information during filtering, which is crucial for further processing, such as angle estimation. Our experiments show, that the use of CVCNNs increases data efficiency, speeds up network training and substantially improves the conservation of phase information during interference removal.
\end{abstract}

\IEEEpeerreviewmaketitle

\section{Introduction}
\label{sec:intro}

Advanced Driver Assistance Systems (ADAS) and Autonomous Vehicles (AV) rely on multi modal sensor data for the perception of the vehicles' surroundings. Radar sensors deliver valuable information about object locations as well as object velocities. Frequency modulated continuous wave (FMCW)/chirp sequence (CS) radars are the most commonly used radar systems for automotive applications. They transmit sequences of linearly modulated chirp signals within a non-regulated spectrum. 

Due to the lack of transmit regulations and the increased traffic volume of autonomous and radar-enhanced vehicles, the chance of mutual interference between multiple radar sensors increases.
In most cases this interference is of non-coherent nature, where the involved radar sensors operate with different transmit parameters.
Non-coherent interference leads to a decreased object detection sensitivity, caused by broadband disturbances within the radar signal.

Typically, interference mitigation of mutual interference is performed using classical signal processing algorithms. The most rudimentary method is to substitute all interference-affected samples with zero~\cite{Fischer}, followed by an optional smoothing of the boundaries. More advanced methods use non-linear filtering in slow-time~\cite{WAG18}, iterative reconstruction using Fourier transforms and thresholding~\cite{MAR12}, estimation and subtraction of the interference component~\cite{BEC17}, or beamforming~\cite{Bechter2016}.

Recently, the use of deep learning and neural networks has been proposed for interference mitigation. Neural networks are typically applied in a supervised manner such that the signal is filtered and the interference is damped. Recurrent Neural Networks (RNNs)~\cite{8690848,9053013} are commonly applied to time-domain signals, while in frequency-domain preferably Convolutional Neural Network (CNN) -based models are used. The proposed CNN-architectures range from very small CNNs~\cite{9114627,raar_rock} over Convolutional Autoencoders~\cite{9114719}, U-Net inspired CNNs~\cite{9114641}, to bigger CNNs that try to learn a signal transformation from STFT to FFT additional to denoising the signal~\cite{Ristea2020}.

\begin{figure}[h]
    \centering
	\begin{minipage}[h]{0.4\columnwidth}
	\vspace{-0.3cm}
	\resizebox{\columnwidth}{!}{
\begin{tikzpicture}

\begin{axis}[
scale only axis=true,
height = 1.4\textwidth,
width = 1.4\textwidth,
axis background/.style={fill=white},
axis line style={white},
colorbar,
colorbar style={ylabel={},width=6,
axis line style = { draw = none },
	yticklabel style={
		text width=width("$-30$"),
		align=right
	}},
colormap/viridis,
point meta max=0,
point meta min=-30,
tick align=outside,
tick pos=left,
x grid style={white},
xlabel={Velocity [m/s]},
xmajorgrids,
xmin=-16, xmax=15.6666666666667,
xtick style={color=white!33.3333333333333!black},
y grid style={white},
ylabel={Distance [m]},
ymajorgrids,
ymin=6, ymax=60,
ytick style={color=white!33.3333333333333!black}
]
\addplot graphics [includegraphics cmd=\pgfimage,xmin=-16, xmax=15.6666666666667, ymin=6, ymax=60] {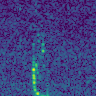};
\end{axis}

\draw ({$(current bounding box.south west)!0.5!(current bounding box.south east)$}|-{$(current bounding box.south west)!1.1!(current bounding box.north west)$}) node[
  scale=1.5,
  anchor=north,
  text=black,
  rotate=0.0
]{(a)};
\end{tikzpicture}}
	\end{minipage}
	\begin{minipage}[h]{0.4\columnwidth}
	\vspace{-0.3cm}
	    \resizebox{\columnwidth}{!}{
\begin{tikzpicture}

\begin{axis}[
scale only axis=true,
height = 1.4\textwidth,
width = 1.4\textwidth,
axis background/.style={fill=white},
axis line style={white},
colorbar,
colorbar style={ylabel={},width=6,axis line style = { draw = none },
	yticklabel style={
		text width=width("$-30$"),
		align=right
	}},
colormap/viridis,
point meta max=0,
point meta min=-30,
tick align=outside,
tick pos=left,
x grid style={white},
xlabel={Velocity [m/s]},
xmajorgrids,
xmin=-16, xmax=15.6666666666667,
xtick style={color=white!33.3333333333333!black},
y grid style={white},
ylabel={Distance [m]},
ymajorgrids,
ymin=6, ymax=60,
ytick style={color=white!33.3333333333333!black}
]
\addplot graphics [includegraphics cmd=\pgfimage,xmin=-16, xmax=15.6666666666667, ymin=6, ymax=60] {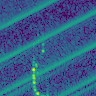};
\end{axis}

\draw ({$(current bounding box.south west)!0.5!(current bounding box.south east)$}|-{$(current bounding box.south west)!1.1!(current bounding box.north west)$}) node[
  scale=1.5,
  anchor=north,
  text=black,
  rotate=0.0
]{(b)};
\end{tikzpicture}}
	\end{minipage}
	\begin{minipage}[h]{0.4\columnwidth}
		\resizebox{\columnwidth}{!}{
\begin{tikzpicture}

\begin{axis}[
scale only axis=true,
height = 1.4\textwidth,
width = 1.4\textwidth,
axis background/.style={fill=white},
axis line style={white},
colorbar,
colorbar style={ylabel={},width=6,axis line style = { draw = none },
	yticklabel style={
		text width=width("$-30$"),
		align=right
	}},
colormap/viridis,
point meta max=0,
point meta min=-30,
tick align=outside,
tick pos=left,
x grid style={white},
xlabel={Velocity [m/s]},
xmajorgrids,
xmin=-16, xmax=15.6666666666667,
xtick style={color=white!33.3333333333333!black},
y grid style={white},
ylabel={Distance [m]},
ymajorgrids,
ymin=6, ymax=60,
ytick style={color=white!33.3333333333333!black}
]
\addplot graphics [includegraphics cmd=\pgfimage,xmin=-16, xmax=15.6666666666667, ymin=6, ymax=60] {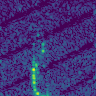};
\end{axis}

\draw ({$(current bounding box.south west)!0.5!(current bounding box.south east)$}|-{$(current bounding box.south west)!1.1!(current bounding box.north west)$}) node[
  scale=1.5,
  anchor=north,
  text=black,
  rotate=0.0
]{(c)};
\end{tikzpicture}}
	\end{minipage}
	\begin{minipage}[h]{0.4\columnwidth}
		\resizebox{\columnwidth}{!}{
\begin{tikzpicture}

\begin{axis}[
scale only axis=true,
height = 1.4\textwidth,
width = 1.4\textwidth,
axis background/.style={fill=white},
axis line style={white},
colorbar,
colorbar style={ylabel={},width=6,
axis line style = { draw = none },
	yticklabel style={
		text width=width("$-30$"),
		align=right
	}},
colormap/viridis,
point meta max=0,
point meta min=-30,
tick align=outside,
tick pos=left,
x grid style={white},
xlabel={Velocity [m/s]},
xmajorgrids,
xmin=-16, xmax=15.6666666666667,
xtick style={color=white!33.3333333333333!black},
y grid style={white},
ylabel={Distance [m]},
ymajorgrids,
ymin=6, ymax=60,
ytick style={color=white!33.3333333333333!black}
]
\addplot graphics [includegraphics cmd=\pgfimage,xmin=-16, xmax=15.6666666666667, ymin=6, ymax=60] {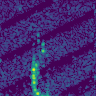};
\end{axis}

\draw ({$(current bounding box.south west)!0.5!(current bounding box.south east)$}|-{$(current bounding box.south west)!1.1!(current bounding box.north west)$}) node[
  scale=1.5,
  anchor=north,
  text=black,
  rotate=0.0
]{(d)};
\end{tikzpicture}}
 	\end{minipage}
	\caption{Range-Doppler map examples; (a) clean data ; (b) with added simulated interference; (c) denoised using a complex-valued CNN; (d) denoised using a real-valued CNN.}
	\label{fig:data_example}
\end{figure}
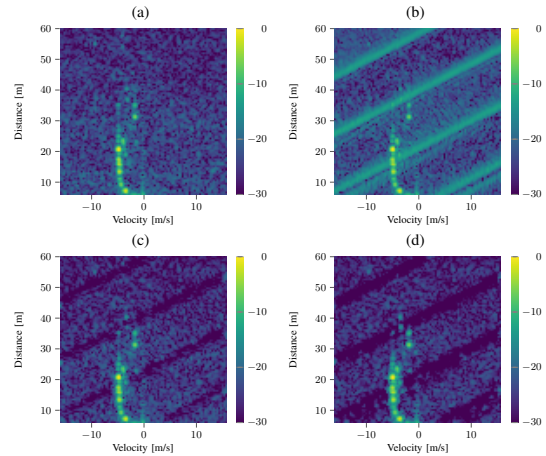

The classical signal processing chain, as depicted in Fig.~\ref{fig:signal_processing_chain}, processes the raw radar signal in several steps to a frequency representation that resembles the characteristics of images. This image-like representation of the signal enables the use of powerful computer-vision methods, e.g CNNs. 
The range-Doppler map, as presented in Fig.~\ref{fig:data_example}, is particularly well suited for denoising and interference mitigation. This complex-valued spectrum is represented as a 2D matrix with signal peaks corresponding to the distances and velocities of objects.
The proposed CNN-based models for interference mitigation substantially improve the results of classical methods when subsequently used for object detection~\cite{9114627,9114641,Ristea2020}. These object detection algorithms are applied to the magnitude spectrum of the interference mitigated RD-maps. However, for phase-dependent tasks none of the approaches (classical or deep learning based methods) are capable of sufficiently reconstructing the phase of the clean signal.
Most of recent works provide the real- and imaginary parts of the complex spectra as inputs, in order to perform phase estimations.
However, they rely purely on data to learn the relationship between real- and imaginary parts of the spectra, adding additional complexity to the learning problem.

Therefore we propose the use of complex-valued CNNs (CVCNNs) for radar signal denoising and interference mitigation. The use of complex-valued analysis introduces an inductive bias, restricting the degrees of freedom in the network. This restriction can however benefit the learning behavior substantially, since it enforces signal transformations according to the physical characteristics of the data and reduces the problem complexity. 

CVCNNs were first proposed in the 1990s~\cite{georgiou1992complex}. Recently, Trabelsi et al.~\cite{DBLP:journals/corr/TrabelsiBSSSMRB17} proposed a CVCNN, that incorporates complex-valued analysis using two real-valued convolution kernels. This enables model training to be performed using conventional backpropagation. CVCNNs have been applied to Synthetic Aperture Radar (SAR) data for semantic segmentation~\cite{8039431} and super-resolution imaging~\cite{8458209} with promising results, however limited in experiments and performance analysis.

In this paper we compare CVCNNs for RD-map denoising and interference mitigation with comparable real-valued networks. We consider particularly small model architectures and present insights regarding parameter efficiency, computational efficiency and data efficiency.

The main contributions of this paper are:
\begin{itemize}
    \item Application of CVCNNs for range-Doppler map denoising and interference mitigation using real-world measurements with simulated interference.
    \item Complexity evaluation in terms of model parameters and operations per sample.
    \item Data efficiency analysis of real- and complex-valued CNNs.
    \item Statistical comparison of CNNs using real- or complex-valued weights with 'state-of-the-art' interference mitigation methods.
\end{itemize}

\section{Signal Model}\label{sec:signal_model}

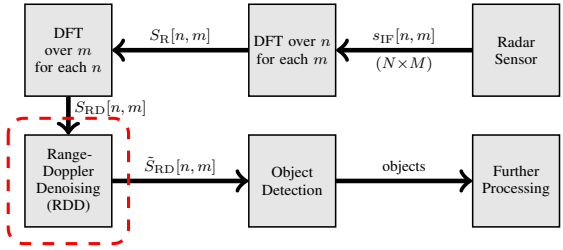
\begin{figure}[tb]
	\centering
	\footnotesize
	\resizebox{8.5cm}{!}{
		\tikzstyle{block}=[draw, fill=black!10, text width=5em, text centered, minimum height=6em]
\tikzstyle{blockfocus}=[block, draw=red, thick, dashed,rounded corners]
\tikzstyle{blockinactive}=[block, opacity=.3]

\tikzstyle{arrow}=[very thick, ->]
\tikzstyle{arrowinactive}=[arrow, opacity=.2]

\begin{tikzpicture}[scale=0.5, transform shape]
\begin{scope}[node distance=0.7cm and 2.5cm]

\node[block] (rs)  at (0,0) {Radar Sensor};
\node[block, left = of rs] (dft1) {DFT over $n$ \\ for each $m$};
\node[block, left = of dft1] (dft2) {DFT over $m$ \\ for each $n$};

\node[block, below = of dft2] (rdd) {Range-Doppler Denoising (RDD)};
\node[block, right = of rdd] (od) {Object Detection};
\node[block, right = of od] (fp) {Further Processing};

\draw[arrow] (rs) -- (dft1) node[midway, above] () {$s_{\mathrm{IF}}[n,m]$} node [midway,below] () {${(N{\times}M)}$};
\draw[arrow] (dft1) -- (dft2) node[midway, above] () {$S_{\mathrm{R}}[n,m]$};
\draw[arrow] (dft2) -- (rdd) node[pos=0.3, right] () {$S_{\mathrm{RD}}[n,m]$};
\draw[arrow] (rdd) -- (od) node[midway, above] () {$\tilde{S}_{\mathrm{RD}}[n,m]$};
\draw[arrow] (od) -- (fp) node[midway, above] () {objects};

\draw[red,thick,dashed,rounded corners] ($(rdd.north west)+(-0.3,0.3)$)  rectangle ($(rdd.south east)+(0.3,-0.3)$);

\end{scope}
\end{tikzpicture}
	}
	\caption{Block diagram of a basic FMCW/CS radar processing chain. The red dashed box indicates the location of the proposed interference mitigation step using a CNN-based approach.}
	\label{fig:signal_processing_chain}
	\vspace{-5mm}
\end{figure}

The \emph{range-Doppler (RD)} processing chain of a common FMCW/CS radar is depicted in Fig.~\ref{fig:signal_processing_chain}. The radar sensor transmits a set of linearly modulated radio frequency (RF) chirps, also termed ramps. Object reflections are perceived by the receive antennas and mixed with the transmit signal resulting in the \emph{Intermediate Frequency (IF) Signal}. The objects' distances and velocities are contained in the sinusoidals' frequencies and their linear phase change over successive ramps~\cite{STO92,WIN07}, respectively.
The signal is processed as an $N \times M$ data matrix $s_{\mathrm{IF}}[n,m]$, containing $N$ \emph{fast time} samples for each of $M$ ramps. Discrete Fourier transforms (DFTs) are computed over both dimensions, yielding a two-dimensional spectrum, the RD-map $S_{\mathrm{RD}}[n,m]$. 
The peaks on the RD-map then correspond to objects' distances and velocities.
After peak detection, further processing can include angular estimation, tracking, and classification.

The IF signal $s_{\mathrm{IF}}[n,m]$ contains object reflections, noise, and may also include interference signals. It is modeled as

\begin{equation}
\small
s_{\mathrm{IF}}[n,m]=\sum_{o=1}^{N_{\mathrm{O}}} s_{\mathrm{O},o}[n,m] + \sum_{i=1}^{N_{\mathrm{I}}} s_{\mathrm{I},i}[n,m] + \upsilon[n,m] \, ,
\label{eq:signal-model}
\end{equation}
where $s_{\mathrm{O},o}[n,m]$ are the $N_{\mathrm{O}}$ object reflections, $s_{\mathrm{I},i}[n,m]$ are interference signals from $N_I$ interfering radars and $\upsilon[n,m]$ models the noise.

State-of-the-art ('classical') interference mitigation methods are mostly signal processing algorithms that are applied either on the time-domain signal $s_{\mathrm{IF}}[n,m]$ or on the frequency-domain signal $S_{\mathrm{R}}[n,m]$ after the first DFT~\cite{TOT18}. The CNN-based method used in this paper, also termed \emph{Range-Doppler Denoising (RDD)\label{rd-denoising}}, is applied on the RD-map after the second DFT (see Fig. \ref{fig:signal_processing_chain}).

\section{Complex-valued convolutional neural networks}\label{sec:complex_cnns}
Complex-valued neural networks are able to solve certain tasks more efficiently than real-valued neural networks~\cite{hirose2012complex}.
Real-Valued CNNs (RVCNNs) that are applied to complex-valued data typically use separate real-valued channels in order to represent the real- and imaginary parts. In this approach, the CNN is required to learn the relation between the real- and imaginary part of the complex-valued data based on the training samples.
In contrast, the CVCNN carries out all operations using complex-valued analysis and therefore incorporates the relation between real- and imaginary parts in the model architecture rather than the learned parameters.
The motivation for this approach comes from the complex-valued nature of the radar signal itself.
In theory, a CVCNN should be capable of processing complex radar spectra more easily than its real-valued alternative which has no intuition about complex numbers.
Like local connectivity in convolution kernels, this introduces an inductive bias to the network architecture.
Therefore, tasks which rely on correct phase relations, as radar signal processing, can greatly benefit from using complex operations within the CNN.
For our proposed architecture we rely on three basic operations: convolutions, batch normalization and an appropriate activation function. All three operations have to be carried out following complex analysis.

\subsection{Complex convolution}\label{sec:complex_convolution}
A convolution kernel for a two-dimensional convolution operation consists of a four dimensional tensor $W_{ijkl}$ of size $k_x\times k_y\times C_{in}\times C_{out}$. Here $k_x$ and $k_y$ represent the spatial size of the kernel, whereas $C_{in}$ indicates the number of input filter channels and $C_{out}$ the number of output filter channels. 
Considering the number of filter channels is a multiple of two, we can always split the channels into two parts, one for the real- and one for the imaginary part. Therefore, the separate kernels have size $k_x\times k_y\times \frac{C_{in}}{2}\times \frac{C_{out}}{2}$.
\begin{figure}[h]
    \centering
    \includegraphics[width=0.9\linewidth]{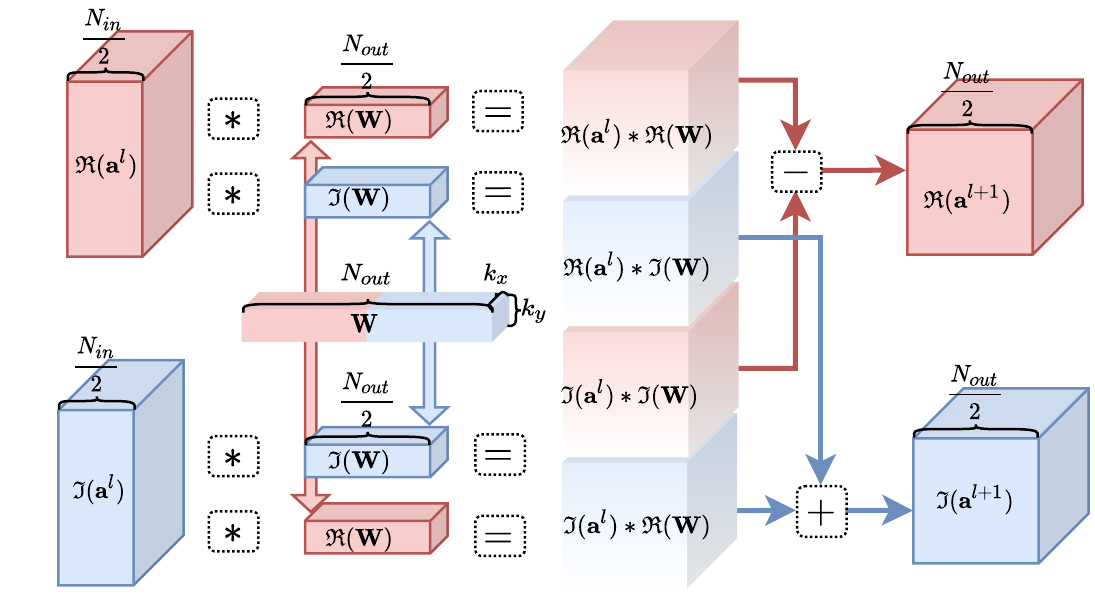}
    \caption{Depiction of a complex convolution on input $\mathbf{a}^l \in \mathbb{C}$ using the kernel $W \in \mathbb{C}$, creating the activation of the next level $\mathbf{a}^{l+1}\in \mathbb{C}$.}
    \label{fig:complex_conv}
\end{figure}
In complex analysis the multiplication of two complex numbers $z_1=(a+ib)$ and $z_2=(x+iy)$ results in $z_1\cdot z_2 = ax-by+i(bx+ay)$,
where $z_1,z_2 \in \mathbb{C}$ and $a,b \in \mathbb{R}$ with $i$ being the imaginary unit. The same works for complex tensors.
Since the convolution operator ($*$) is distributive we obtain, 
\begin{equation}
\small
	\mathbf{W} * \mathbf{h} = (\mathbf{A}+i\mathbf{B}) * (\mathbf{x}+i\mathbf{y})= \mathbf{A}*\mathbf{x}-\mathbf{B}*\mathbf{y}+i(\mathbf{B}*\mathbf{x}+\mathbf{A}*\mathbf{y}),
\end{equation} 
convolving the complex vector $\mathbf{h}= \mathbf{x}+i\mathbf{y}$ with the complex kernel matrix $\mathbf{W} = \mathbf{A}+i\mathbf{B}$~\cite{DBLP:journals/corr/TrabelsiBSSSMRB17}. Therefore, the complex-valued convolution can be executed as a series of real-valued convolutions, as shown in Fig.~\ref{fig:complex_conv}.

\subsection{Complex Batch Normalization}\label{sec:complex_bn}
Deep CNNs rely on the Batch Normalization (BN) operation to normalize activations and accelerate training~\cite{DBLP:journals/corr/IoffeS15}. Since the standard formulation of BN only applies to real-valued activations, complex-valued networks require different methods.
One possibility to realize BN for complex numbers was proposed in~\cite{DBLP:journals/corr/TrabelsiBSSSMRB17}, where they use a whitening procedure to standardize the complex numbers. First the inputs $\mathbf{z} \in \mathbb{C}$ are whitened and then standard BN is performed. Thus, the vector after the whitening step is scaled and shifted using $\boldsymbol{\gamma}$ and $\boldsymbol{\beta}$ respectively.
\begin{equation}
\small
	\mathrm{BN}(\mathbf{z})= \boldsymbol{\gamma}\odot (\mathbf{V})^{-\frac{1}{2}} (\mathbf{z} - \mathbb{E}[\mathbf{z}]) +\boldsymbol{\beta},
\end{equation}
where $\mathbf{V}$ is a $2\times 2$ positive (semi)-definite \emph{covariance matrix}. 
Since the proposed BN needs the square root of the inverse of $\mathbf{V}$, positive definiteness of the matrix is ensured by adding $\epsilon \mathbf{I}$ to $V$, which is known as Tikhonov  regularization.

\subsection{Complex Rectified Linear Unit ($\mathbb{C}\mathrm{ReLU}$)}\label{sec:complex_act_fn}

The complex-valued ReLU function is defined as
\begin{equation}
\small
	\mathbb{C}\mathrm{ReLU}(z)= \mathrm{ReLU}(\Re (z))+ i\mathrm{ReLU}(\Im(z)).
\end{equation}
Although it does not fulfil the Cauchy-Riemann equations everywhere and is therefore only holomorphic on a subset of $\mathbb{C}$, but it can be computed efficiently and experiments have shown superior performance compared to other proposed complex-valued activation functions~\cite{DBLP:journals/corr/TrabelsiBSSSMRB17}.

\section{CVCNN for range-Doppler processing}\label{sec:complex_cnn_architectures}
Our CVCNN takes RD-maps with interference as inputs and predicts denoised RD-maps as outputs. 

\subsection{Model}\label{sec:model}

\begin{figure}
	\centering
	\footnotesize
	\includegraphics[width=0.82\columnwidth]{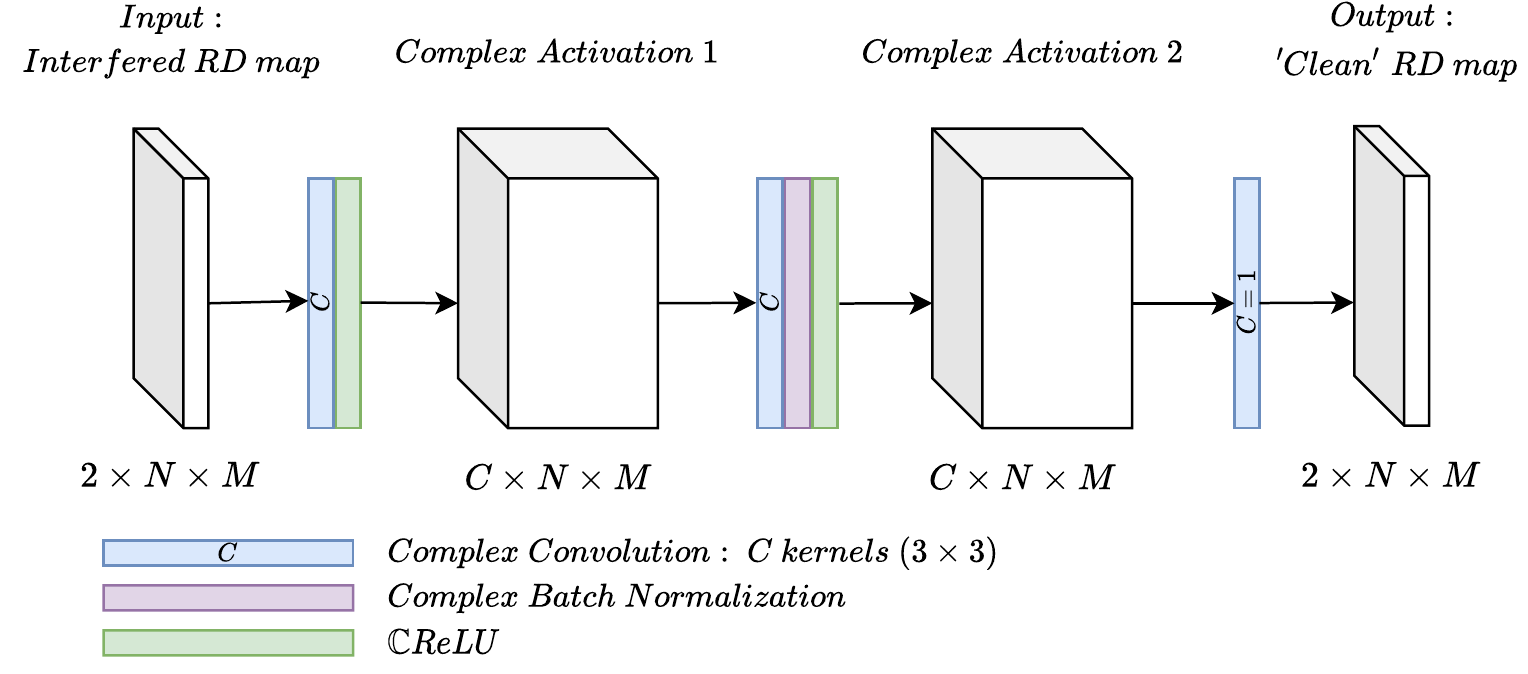}
	\caption{CVCNN architecture for radar signal denoising and interference mitigation. It uses $\mathbb{C}\mathrm{ReLU}$, complex BN and the complex-convolution.}
	\label{fig:cnn_arch}
\end{figure}

The proposed CVCNN architecture for interference mitigation and denoising of radar signals is shown in Fig. \ref{fig:cnn_arch} and based on the real-valued approach described in~\cite{Rock1907:Complex}.
The CNN input is a noisy range-Doppler map, which is a complex-valued two dimensional matrix. This complex-valued matrix can be represented as a three dimensional tensor of size $C\times N \times M$ with $N$ and $M$ being the height and width of the RD-maps and $C=2$ the real- and imaginary parts of the complex spectra.
The model architecture implements a fully convolutional NN; it consists exclusively of convolutions, BN operations and the ReLU activation function.
The convolution strides are set to one and 'same' zero padding is used such that the individual feature maps maintain a constant spatial size between layers.
The CVCNN carries out all operations according to complex analysis as described in Section~\ref{sec:complex_cnns}.
Therefore, three layers are constructed as follows:
\begin{enumerate}
	\item The first layer performs a complex-convolution followed by a $\mathbb{C}\mathrm{ReLU}$ non-linearity.
	\item In the second layer the complex-convolution is followed by a complex BN and a $\mathbb{C}\mathrm{ReLU}$ non-linearity.
	\item The last layer consists solely of the convolution operation to create the network output.
\end{enumerate}
\subsection{Relation of computational and parameter complexity}\label{sec:computational_complexity}
\begin{figure}[h]
	\centering
	\resizebox {0.8\linewidth} {!} {
\begin{tikzpicture}

\definecolor{color0}{rgb}{0.886274509803922,0.290196078431373,0.2}
\definecolor{color1}{rgb}{0.203921568627451,0.541176470588235,0.741176470588235}

\begin{axis}[
scale only axis=true,
width=0.9\linewidth,
height = 0.4\linewidth,
axis line style={white},
legend cell align={left},
legend style={fill opacity=0.8, draw opacity=1, text opacity=1, at={(0.03,0.97)}, anchor=north west, draw=white},
tick align=outside,
tick pos=left,
xlabel={Parameters},
xmajorgrids,
xmin=0, xmax=3100,
xtick style={color=white!33.3333333333333!black},
ylabel={MFLOP/RD-map},
ymajorgrids,
ymin=0, ymax=120.0,
ytick style={color=white!33.3333333333333!black}
]
\addplot [semithick, color0, mark=*, mark size=3, mark options={solid}, only marks]
table {%
622 11.300051
954 17.291107
1618 29.273219
2946 53.237443
};
\addlegendentry{Real-valued networks}
\addplot [semithick, color1, mark=*, mark size=3, mark options={solid}, only marks]
table {%
490 17.602827
834 29.878729
1522 54.430533
2898 103.534141
};
\addlegendentry{Complex-valued networks}
\draw (axis cs:650,2.300051) node[
  scale=1.0,
  anchor=base east,
  text=black,
  rotate=0.0
]{16-2-2};
\draw (axis cs:1200,5.0) node[
  scale=1.0,
  anchor=base east,
  text=black,
  rotate=0.0
]{16-4-2};
\draw (axis cs:1600,29.273219) node[
  scale=1.0,
  anchor=base east,
  text=black,
  rotate=0.0
]{16-8-2};
\draw (axis cs:2946,53.237443) node[
  scale=1.0,
  anchor=base east,
  text=black,
  rotate=0.0
]{16-16-2};
\draw (axis cs:450,17.602827) node[
  scale=1.0,
  anchor=base east,
  text=black,
  rotate=0.0
]{8-2-1};
\draw (axis cs:834,29.878729) node[
  scale=1.0,
  anchor=base east,
  text=black,
  rotate=0.0
]{8-4-1};
\draw (axis cs:1522,54.430533) node[
  scale=1.0,
  anchor=base east,
  text=black,
  rotate=0.0
]{8-8-1};
\draw (axis cs:2898,103.534141) node[
  scale=1.0,
  anchor=base east,
  text=black,
  rotate=0.0
]{8-16-1};
\end{axis}

\end{tikzpicture}}
	\caption{Relation of the computational complexity and the number of model parameters. The labels of the networks indicate the number of filter channels in the corresponding layers.}\label{fig:flops_vs_params}
\end{figure}
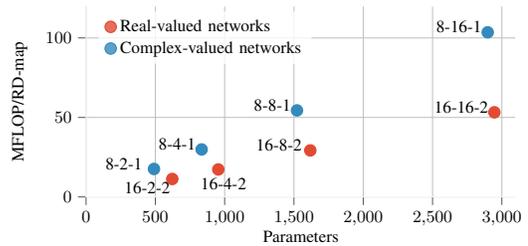

Since CVCNNs have a higher computational complexity per-parameter, we present the number of mega (=million) floating point operations per RD-map (MFLOP/RD-map) over the number of parameters in Fig.~\ref{fig:flops_vs_params}. Note, that we consider the computational complexity per RD-sample for one prediction step of the network. As expected, we see that there is a linear relationship between the number of used filter channels and the computational complexity for both real-valued and complex-valued models. However, the linear relation of the complex-valued models has a larger slope than for the real-valued models. This means that a complex-valued model using the same number of parameters has a higher computational complexity than its real-valued counterpart.

\section{Experimental setup}
We use real-world FMCW/CS radar measurements combined with simulated interference according to \eqref{eq:signal-model}. The signals with and without interference are used as input-output pairs for training the CNN models in order to perform denoising and interference mitigation. The model is applied to the processed radar signal after the second DFT, i.e. the RD-map. We aim to correctly detect peaks in the RD-map that correspond to real objects rather than clutter or noise.

\subsection{Data set}
The measurements were recorded in typical inner-city scenarios, where each measurement consists of 32 consecutive radar snapshots (RD-maps) each captured with sixteen antennas. The radar signal contains reflections from static and moving objects as well as receiver noise. The simulated interference, that is added to the time-domain measurement signal, is generated by sampling uniformly from the ego radar and interferer radar transmit parameters. See~\cite{Rock1907:Complex} for a detailed description of the simulation parameters and~\cite{9114627, toth2020analysis} for an extensive analysis of the used measurement signals.

\subsection{Evaluation metrics}\label{sec:eval_metrics}
The models are evaluated using the F1-Score, Error Vector Magnitude (EVM) and Peak Phase Mean Squared Error (PPMSE), defined as follows:

\subsubsection{F1-Score}\label{sec:f1_score}
The F1-Score is a combined classification measure including the true-positive $t_p$, false-positive $f_p$ and false-negative $f_n$ object detections. It is defined as
\begin{equation}
\small
F_1 = 2 \cdot \dfrac{t_p}{t_p+\frac{1}{2}(f_p+f_n)}.
\end{equation}

For the F1-Score calculation, we first use the Cell Averaging - Constant False Alarm Rate (CA-CFAR) peak detector on the RD-map prediction $S_{\mathrm{RD}}[n,m]$ in order to create a binary object detection map. This object detection map forms, in combination with the ground truth object detection map, the basis for calculating $t_p$, $f_p$ and $f_n$.

\subsubsection{Error Vector Magnitude (EVM) }\label{sec:evm}
The EVM measures the deviation of the predicted RD-map $S_{\mathrm{RD}}[n,m]$ from the clean RD-map $S_{\mathrm{RD,clean}}[n,m]$ at ground truth peak locations. It it given as
\begin{equation}
\small
\mathrm{EVM}= \frac{1}{N_\mathrm{O}} \sum_{\{n,m\}\in \mathcal{O}}\dfrac{|S_{\mathrm{RD,clean}}[n,m]-S_{\mathrm{RD}}[n,m]|}{|S_{\mathrm{RD,clean}}[n,m]|},
\end{equation}
where $n$ and $m$ are the indices of the object, within the set of cells $\mathcal{O}$, containing object peaks, and $N_\mathrm{O}$ is the number of detected peaks.

\subsubsection{Peak Phase Mean Squared Error (PPMSE)}\label{sec:ppmse}
The PPMSE measures the average squared difference between the angle of the clean range-Doppler signal vector $S_{\mathrm{RD,clean}}[n,m]$ and the predicted range-Doppler signal vector $S_{\mathrm{RD}}[n,m]$ of all detected object peaks. It is defined as
\begin{equation}
\small
\begin{split}
&\Delta[n,m] = \left\lvert\mathrm{atan2}\left(\frac{\Im(S_{\mathrm{RD}}[n,m])}{\Re(S_{\mathrm{RD}}[n,m])}\right)-\mathrm{atan2}\left(\frac{\Im(S_{\mathrm{RD,clean}}[n,m])}{\Re(S_{\mathrm{RD,clean}}[n,m])}\right)\right\rvert,\\
&\mathrm{PPMSE}[n,m] = \frac{1}{N_\mathrm{O}} \sum_{\{n,m\}\in \mathcal{O}}\min (\Delta[n,m],2 \pi-\Delta[n,m])^2,
\end{split}
\end{equation}
where $n$ and $m$ are the indices of cells containing object peaks $\mathcal{O}$ and $N_\mathrm{O}$ is the number of detected peaks. 
\subsection{Training settings}\label{sec:training_settings}
The interfered RD-maps $S_{\mathrm{RD,interfered}}$ are used as inputs for the CNN, while the clean RD-maps $S_{\mathrm{RD,clean}}$ are used as the training targets. All RD-maps are cropped to $N\times M = 96\ \textrm{range cells} \times 96\ \textrm{Doppler cells}$ and subsequently scaled to zero-mean and unit-variance.
The models were trained for 100 epochs using Adam \cite{DBLP:journals/corr/KingmaB14} with a learning rate of $5\cdot 10^{-3}$ and a mini-batch size of $L=8$. As our training objective, we use the mean squared error (MSE) of real- and imaginary parts of the predictions $S_{\mathrm{RD}}$ and targets $S_{\mathrm{RD,clean}}$

\section{Experiments}\label{sec:experiments}
We evaluate the denoising capabilities of CVCNNs for RD-maps with simulated interference. We compare complex-valued networks with their real-valued equivalent evaluating the computational, parameter and data efficiency. Furthermore, we analyze the performance of both CNN approaches against classical methods.   

\subsection{Parameter efficiency}\label{sec:parameter_efficiency}
In this experiments we investigate the parameter efficiency of CVCNNs compared to their real-valued counterparts. Therefore, we train and evaluate 24 different architectures varying the number of filter channels $C_i$ per layer $i$.
For the complex-valued network we vary the number of (complex-valued) filter kernels (= channels in the next activation map) in the first layer using $C_1 \in \{1,2,4,8,16,32\}$ kernels and in the second layer using $C_2 \in \{2,4,8,16\}$ kernels. The output layer is fixed to $C_3 = 1$ complex-valued filter kernel.
The real-valued network uses $C_1 \in \{1,2,4,8,16,32\}$, $C_2 \in \{4,8,16,32\}$ and $C_3 = 2$ real-valued filter kernels. We combine the sets of kernels per layer as a Cartesian product and thus evaluate all possible combinations, yielding 24 architectures for each approach.

\begin{figure}[h]
    {
    \small
	\begin{minipage}[h]{0.9\columnwidth}
	    \hspace{0.05\textwidth}
		\resizebox {\textwidth} {!} {
            \begin{tikzpicture}

\definecolor{color0}{rgb}{0.886274509803922,0.290196078431373,0.2}
\definecolor{color1}{rgb}{0.203921568627451,0.541176470588235,0.741176470588235}
\definecolor{color2}{rgb}{0.976,0.859,0.839}
\definecolor{interferedorange}{rgb}{1,0.498039215686275,0.0549019607843137}
\definecolor{cleangreen}{rgb}{0.172549019607843,0.627450980392157,0.172549019607843}

\centering
\begin{axis}[%
hide axis,
xmin=0,
xmax=1,
ymin=0,
ymax=1,
scale=0.5,
legend columns=4,
legend style={fill opacity=0.8, draw opacity=1, text opacity=1, draw=white!80!black},
]
\addlegendimage{color0,mark=*}
\addlegendentry{Real-valued models}
\addlegendimage{color1,mark=*}
\addlegendentry{Complex-valued models}
\addlegendimage{cleangreen,dashed, line width=2pt}
\addlegendentry{Target}
\addlegendimage{interferedorange, dashed, line width=2pt}
\addlegendentry{No Mitigation}
\end{axis}
\end{tikzpicture}}
            \vspace{-1.8cm}
	\end{minipage}
	\begin{minipage}[h]{0.49\columnwidth}
		\resizebox {\textwidth} {!} {
\begin{tikzpicture}

\definecolor{color0}{rgb}{0.886274509803922,0.290196078431373,0.2}
\definecolor{color1}{rgb}{1,0.812,0.282}
\definecolor{color2}{rgb}{0.203921568627451,0.541176470588235,0.741176470588235}
\definecolor{color3}{rgb}{0.11,0.827,0.635}
\definecolor{interferedorange}{rgb}{1,0.498039215686275,0.0549019607843137}
\definecolor{cleangreen}{rgb}{0.172549019607843,0.627450980392157,0.172549019607843}

\pgfdeclarelayer{fg}    
\pgfsetlayers{main,fg}  

\begin{axis}[
scale only axis=true,
height=0.7\columnwidth,
width=1.0\columnwidth,
axis line style={white},
ticklabel style={
/pgf/number format/fixed,
/pgf/number format/precision=1
},
scaled ticks = false,
legend cell align={left},
legend style={fill opacity=0.8, draw opacity=1, text opacity=1, draw=white!80!black},
tick align=outside,
tick pos=left,
title={\textbf{F1-Score}},
xlabel={Parameters},
xmajorgrids,
xmin=-426, xmax=10750,
xtick style={color=white!33.3333333333333!black},
ymajorgrids,
ymin=0.8, ymax=0.92,
ytick style={color=white!33.3333333333333!black}
]
\addplot [semithick, color0, mark=*, mark size=2, mark options={solid}, only marks]
table {%
82 0.743411958217621
144 0.85270893573761
268 0.811602890491486
516 0.854615807533264
118 0.836319148540497
198 0.875165700912476
358 0.884819865226746
678 0.896835565567017
190 0.873671412467957
306 0.89341789484024
538 0.887760698795319
1002 0.895070374011993
334 0.868516325950623
522 0.893086612224579
898 0.875819563865662
1650 0.877782642841339
622 0.875872135162354
954 0.891342878341675
1618 0.88917475938797
2946 0.896046757698059
1198 0.884306788444519
1818 0.889905869960785
3058 0.888761937618256
5538 0.896435022354126
};
\addplot [semithick, color2, mark=*, mark size=2, mark options={solid}, only marks]
table {%
112 0.879345595836639
204 0.8802410364151
388 0.894991636276245
756 0.895353734493256
166 0.887373387813568
294 0.891399145126343
550 0.899133682250977
1062 0.899132966995239
274 0.894851803779602
474 0.898428201675415
874 0.900455594062805
1674 0.902918696403503
490 0.899037480354309
834 0.89675360918045
1522 0.899232387542725
2898 0.901084184646606
922 0.893654584884644
1554 0.897091686725616
2818 0.900775492191315
5346 0.901577234268188
1786 0.896572232246399
2994 0.89754182100296
5410 0.899857640266418
10242 0.901313662528992
};
\addplot [semithick, color1, mark=*, mark size=2, mark options={solid}, only marks]
table {%
678 0.896835565567017
};
\addplot [semithick, color3, mark=*, mark size=2, mark options={solid}, only marks]
table {%
1674 0.902918696403503
};
\draw (axis cs:3100,0.90) node[
  scale=0.6,
  anchor=base west,
  text=black,
  rotate=0.0
]{};
\draw (axis cs:1200,0.8) node[
  scale=0.6,
  anchor=base west,
  text=black,
  rotate=0.0
]{};
\draw (axis cs:678,0.90) node(R){};
\draw (axis cs:2600,0.87) node (RR){};
\draw (axis cs:1674,0.903) node (C){} ;
\draw (axis cs:5000,0.91) node (CC){};

\addplot [dashed, line width=2pt, cleangreen]
table {%
0 0.908827448042987
10000 0.908827448042987
};
\addplot [dashed, line width=2pt, interferedorange]
table {%
0 0.836633657524875
10000 0.836633657524875
};

\end{axis}

\begin{pgfonlayer}{fg}    
\draw[<-] (R) -- (RR) node[below,scale=0.8] () {Best-$\mathbb{R}$@678};
\draw[<-] (C) -- (CC) node[above,scale=0.8] () {Best-$\mathbb{C}$@1674};
\end{pgfonlayer}

\end{tikzpicture}}
	\end{minipage}%
	\begin{minipage}[h]{0.49\columnwidth}
			\resizebox {\textwidth} {!} {
\begin{tikzpicture}

\definecolor{color0}{rgb}{0.886274509803922,0.290196078431373,0.2}
\definecolor{color1}{rgb}{1,0.812,0.282}
\definecolor{color2}{rgb}{0.203921568627451,0.541176470588235,0.741176470588235}
\definecolor{color3}{rgb}{0.11,0.827,0.635}
\definecolor{interferedorange}{rgb}{1,0.498039215686275,0.0549019607843137}
\definecolor{cleangreen}{rgb}{0.172549019607843,0.627450980392157,0.172549019607843}

\pgfdeclarelayer{fg}    
\pgfsetlayers{main,fg}  

\begin{axis}[
scale only axis=true,
height=0.7\columnwidth,
width=1.0\columnwidth,
axis line style={white},
ticklabel style={
/pgf/number format/fixed,
/pgf/number format/precision=1
},
scaled ticks = false,
legend cell align={left},
legend style={fill opacity=0.8, draw opacity=1, text opacity=1, draw=white!80!black, fill=white!89.8039215686275!black},
tick align=outside,
tick pos=left,
title={\textbf{EVM}},
xlabel={Parameters},
xmajorgrids,
xmin=-426, xmax=10750,
xtick style={color=white!33.3333333333333!black},
ymajorgrids,
ymin=0.0, ymax=0.5,
ytick style={color=white!33.3333333333333!black}
]
\addplot [semithick, color0, mark=*, mark size=2, mark options={solid}, only marks]
table {%
82 0.606750667095184
144 0.504126906394958
268 0.509726285934448
516 0.46876847743988
118 0.338867634534836
198 0.309019953012466
358 0.251329392194748
678 0.220488011837006
190 0.214063048362732
306 0.214092791080475
538 0.185415983200073
1002 0.158666536211967
334 0.230401396751404
522 0.181352153420448
898 0.159862145781517
1650 0.143454611301422
622 0.203363195061684
954 0.191626161336899
1618 0.14835724234581
2946 0.122640661895275
1198 0.20906688272953
1818 0.155429318547249
3058 0.134613364934921
5538 0.153856962919235
};
\addplot [semithick, color2, mark=*, mark size=2, mark options={solid}, only marks]
table {%
112 0.331424355506897
204 0.294220924377441
388 0.239244729280472
756 0.212833315134048
166 0.230952858924866
294 0.203660979866982
550 0.179849922657013
1062 0.153447866439819
274 0.165928244590759
474 0.142029225826263
874 0.129205524921417
1674 0.116188637912273
490 0.135618284344673
834 0.121494293212891
1522 0.119222089648247
2898 0.110702328383923
922 0.117932714521885
1554 0.107008084654808
2818 0.107367962598801
5346 0.102251350879669
1786 0.118278115987778
2994 0.106835573911667
5410 0.105434842407703
10242 0.0981575548648834
};

\addplot [semithick, color1, mark=*, mark size=2, mark options={solid}, only marks]
table {%
2946 0.122640661895275
};

\addplot [semithick,color3, mark=*, mark size=2, mark options={solid}, only marks]
table {%
10242 0.0981575548648834
};

\draw (axis cs:3200,0.13) node[
  scale=0.6,
  anchor=base west,
  text=black,
  rotate=0.0
]{};

\draw (axis cs:8000,0.102144464850426) node[
  scale=0.6,
  anchor=base west,
  text=black,
  rotate=0.0
]{};
\draw (axis cs:2946,0.122640661895275) node(R){};
\draw (axis cs:3000,0.3) node (RR){};
\draw (axis cs:10242,0.0981575548648834) node (C){} ;
\draw (axis cs:8000,0.2) node (CC){};

\addplot [dashed, line width=2pt, cleangreen]
table {%
0 0
10000 0
};
\addplot [dashed, line width=2pt, interferedorange]
table {%
0 0.08736349517682467
10000 0.08736349517682467
};

\end{axis}

\begin{pgfonlayer}{fg}    
\draw[<-] (R) -- (RR) node[above,scale=0.8] () {Best-$\mathbb{R}$@2946};
\draw[<-] (C) -- (CC) node[above,scale=0.8] () {Best-$\mathbb{C}$@10242};
\end{pgfonlayer}

\end{tikzpicture}}
	\end{minipage}
	}
	\caption{Parameter efficiency of RVCNNs and CVCNNs. F1-Score and EVM over the number of parameters were used as performance metrics. The best real-valued and complex-valued models are annotated with their number of parameters.}
	\label{fig:rd-denoising_param}
\end{figure}
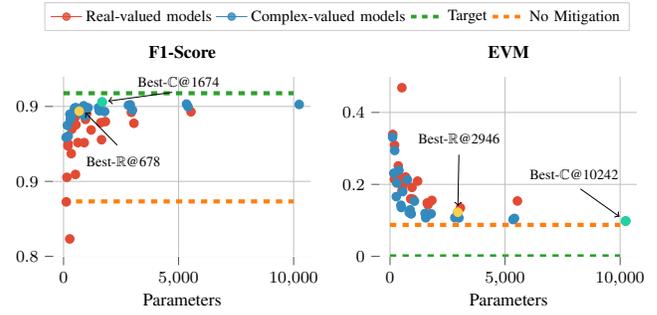
The results in Fig.~\ref{fig:rd-denoising_param} show, that the complex-valued networks outperform the real-valued networks with respect to performance per parameter count. All metrics show substantial improvements across all architecture combinations. Complex-valued network architectures below 500 parameters can deliver comparable performance to even the largest used real-valued networks, reducing the overall memory footprint of the network. 

\subsection{Computational efficiency}\label{sec:computational_efficiency}
Parameter efficiency does not directly translate to computational efficiency, since complex-valued operations are slightly more computationally demanding than real-valued operations as discussed in Section~\ref{sec:computational_complexity}. Therefore, the same architectures as above were used in order to evaluate the performance of the models w.r.t their computational efficiency in MFLOP/RD-map. Note, that we consider the computational complexity per RD-sample for one prediction step of the network.
\begin{figure}[h]
    {
    \small
	\begin{minipage}[h]{0.9\columnwidth}
	    \hspace{0.05\textwidth}
		\resizebox {\textwidth} {!} {
            \begin{tikzpicture}

\definecolor{color0}{rgb}{0.886274509803922,0.290196078431373,0.2}
\definecolor{color1}{rgb}{0.203921568627451,0.541176470588235,0.741176470588235}
\definecolor{color2}{rgb}{0.976,0.859,0.839}
\definecolor{interferedorange}{rgb}{1,0.498039215686275,0.0549019607843137}
\definecolor{cleangreen}{rgb}{0.172549019607843,0.627450980392157,0.172549019607843}

\centering
\begin{axis}[%
hide axis,
xmin=0,
xmax=1,
ymin=0,
ymax=1,
scale=0.5,
legend columns=4,
legend style={fill opacity=0.8, draw opacity=1, text opacity=1, draw=white!80!black},
]
\addlegendimage{color0,mark=*}
\addlegendentry{Real-valued models}
\addlegendimage{color1,mark=*}
\addlegendentry{Complex-valued models}
\addlegendimage{cleangreen,dashed, line width=2pt}
\addlegendentry{Target}
\addlegendimage{interferedorange, dashed, line width=2pt}
\addlegendentry{No Mitigation}
\end{axis}
\end{tikzpicture}}
            \vspace{-1.8cm}
	\end{minipage}
	\begin{minipage}[h]{0.49\columnwidth}
			\resizebox {\textwidth} {!} {
\begin{tikzpicture}[spy using outlines= {circle, magnification=1.5, connect spies}]


\definecolor{color0}{rgb}{0.886274509803922,0.290196078431373,0.2}
\definecolor{color1}{rgb}{1,0.812,0.282}
\definecolor{color2}{rgb}{0.203921568627451,0.541176470588235,0.741176470588235}
\definecolor{color3}{rgb}{0.11,0.827,0.635}
\definecolor{interferedorange}{rgb}{1,0.498039215686275,0.0549019607843137}
\definecolor{cleangreen}{rgb}{0.172549019607843,0.627450980392157,0.172549019607843}

\pgfdeclarelayer{fg}    
\pgfsetlayers{main,fg}  

\begin{axis}[
scale only axis=true,
height=0.7\columnwidth,
width=1.0\columnwidth,
axis line style={white},
legend cell align={left},
legend style={fill opacity=0.8, draw opacity=1, text opacity=1, draw=white!80!black, fill=white!89.8039215686275!black},
tick align=outside,
tick pos=left,
title={\textbf{F1-Score}},
xlabel={MFLOP/RD-map},
xmajorgrids,
xmin=-17.3003759, xmax=392.9130959,
xtick style={color=white!33.3333333333333!black},
ymajorgrids,
ymin=0.8, ymax=0.92,
ytick style={color=white!33.3333333333333!black}
]
\addplot [semithick, color0, mark=*, mark size=2, mark options={solid}, only marks]
table {%
1.345691 0.743411958217621
2.359567 0.85270893573761
4.387319 0.811602890491486
8.442823 0.854615807533264
2.009315 0.836319148540497
3.355003 0.875165700912476
6.046379 0.884819865226746
11.429131 0.896835565567017
3.336563 0.873671412467957
5.345875 0.89341789484024
9.364499 0.887760698795319
17.401747 0.895070374011993
5.991059 0.868516325950623
9.327619 0.893086612224579
16.000739 0.875819563865662
29.346979 0.877782642841339
11.300051 0.875872135162354
17.291107 0.891342878341675
29.273219 0.88917475938797
53.237443 0.896046757698059
21.918035 0.884306788444519
33.218083 0.889905869960785
55.818179 0.888761937618256
101.018371 0.896435022354126
};

\addplot [semithick, color2, mark=*, mark size=2, mark options={solid}, only marks]
table {%
3.668046 0.883072376251221
6.654094 0.878831267356873
12.62619 0.89515084028244
24.570382 0.894320666790009
5.658729 0.888959348201752
9.971899 0.895028054714203
18.598239 0.899849236011505
35.850919 0.90104466676712
9.640095 0.895117580890656
16.607509 0.899990975856781
30.542337 0.900842785835266
58.411993 0.902637839317322
17.602827 0.895883560180664
29.878729 0.898633241653442
54.430533 0.897458791732788
103.534141 0.901319026947021
33.528291 0.890114665031433
56.421169 0.895442128181458
102.206925 0.900367856025696
193.778437 0.900267481803894
65.379219 0.891593396663666
109.506049 0.893929600715637
197.759709 0.899325668811798
374.267029 0.901998281478882
};
\addplot [semithick, color1, mark=*, mark size=2, mark options={solid}, only marks]
table {%
11.429131 0.896835565567017
};
\addplot [semithick, color3, mark=*, mark size=2, mark options={solid}, only marks]
table {%
58.411993 0.902637839317322
};

\draw (axis cs:11.429131,0.896835565567017) node (R)[]{};
\draw (axis cs:150,0.88) node (RR){};
\draw(axis cs:58.411993,0.902637839317322) node (C) []{};
\draw (axis cs:150,0.915) node (CC){};

\begin{pgfonlayer}{fg}    
\draw[<-] (R) -- (RR) node[below,scale=0.8] () {Best-$\mathbb{R}$@11};
\draw[<-] (C) -- (CC) node[right,scale=0.8] () {Best-$\mathbb{C}$@58};
\end{pgfonlayer}

\addplot [dashed, line width=2pt, cleangreen]
table {%
0 0.908827448042987
400 0.908827448042987
};
\addplot [dashed, line width=2pt, interferedorange]
table {%
0 0.836633657524875
400 0.836633657524875
};

\end{axis}


\end{tikzpicture}}
	\end{minipage}%
	\begin{minipage}[h]{0.49\columnwidth}
		\resizebox {\textwidth} {!} {
\begin{tikzpicture}

\definecolor{color0}{rgb}{0.886274509803922,0.290196078431373,0.2}
\definecolor{color1}{rgb}{1,0.812,0.282}
\definecolor{color2}{rgb}{0.203921568627451,0.541176470588235,0.741176470588235}
\definecolor{color3}{rgb}{0.11,0.827,0.635}
\definecolor{interferedorange}{rgb}{1,0.498039215686275,0.0549019607843137}
\definecolor{cleangreen}{rgb}{0.172549019607843,0.627450980392157,0.172549019607843}

\pgfdeclarelayer{fg}    
\pgfsetlayers{main,fg}  

\begin{axis}[
scale only axis=true,
height=0.7\columnwidth,
width=1.0\columnwidth,
axis line style={white},
legend cell align={left},
legend style={fill opacity=0.8, draw opacity=1, text opacity=1, draw=white!80!black, fill=white!89.8039215686275!black},
tick align=outside,
tick pos=left,
title={\textbf{EVM}},
xlabel={MFLOP/RD-map},
xmajorgrids,
xmin=-17.3003759, xmax=392.9130959,
xtick style={color=white!33.3333333333333!black},
ymajorgrids,
ymin=0.0, ymax=0.5,
ytick style={color=white!33.3333333333333!black}
]
\addplot [semithick, color0, mark=*, mark size=2, mark options={solid}, only marks]
table {%
1.345691 0.606750667095184
2.359567 0.504126906394958
4.387319 0.509726285934448
8.442823 0.46876847743988
2.009315 0.338867634534836
3.355003 0.309019953012466
6.046379 0.251329392194748
11.429131 0.220488011837006
3.336563 0.214063048362732
5.345875 0.214092791080475
9.364499 0.185415983200073
17.401747 0.158666536211967
5.991059 0.230401396751404
9.327619 0.181352153420448
16.000739 0.159862145781517
29.346979 0.143454611301422
11.300051 0.203363195061684
17.291107 0.191626161336899
29.273219 0.14835724234581
53.237443 0.122640661895275
21.918035 0.20906688272953
33.218083 0.155429318547249
55.818179 0.134613364934921
101.018371 0.153856962919235
};
\addplot [semithick, color2, mark=*, mark size=2, mark options={solid}, only marks]
table {%
3.668046 0.331424355506897
6.654094 0.294220924377441
12.62619 0.239244729280472
24.570382 0.212833315134048
5.658729 0.230952858924866
9.971899 0.203660979866982
18.598239 0.179849922657013
35.850919 0.153447866439819
9.640095 0.165928244590759
16.607509 0.142029225826263
30.542337 0.129205524921417
58.411993 0.116188637912273
17.602827 0.135618284344673
29.878729 0.121494293212891
54.430533 0.119222089648247
103.534141 0.110702328383923
33.528291 0.117932714521885
56.421169 0.107008084654808
102.206925 0.107367962598801
193.778437 0.102251350879669
65.379219 0.118278115987778
109.506049 0.106835573911667
197.759709 0.105434842407703
374.267029 0.0981575548648834
};
\addplot [semithick, color1, mark=*, mark size=2, mark options={solid}, only marks]
table {%
53.237443 0.122640661895275
};
\addplot [semithick, color3, mark=*, mark size=2, mark options={solid}, only marks]
table {%
374.267029 0.0981575548648834
};
\draw (axis cs:53.2,0.122) node(R){};
\draw (axis cs:100,0.3) node (RR){};
\draw (axis cs:374.267029,0.0981575548648834) node (C){} ;
\draw (axis cs:250,0.2) node (CC){};

\addplot [dashed, line width=2pt, cleangreen]
table {%
0 0
400 0
};
\addplot [dashed,  line width=2pt, interferedorange]
table {%
0 0.08736349517682467
400 0.08736349517682467
};

\end{axis}

\begin{pgfonlayer}{fg}    
\draw[<-] (R) -- (RR) node[above,scale=0.8] () {Best-$\mathbb{R}$@53};
\draw[<-] (C) -- (CC) node[above,scale=0.8] () {Best-$\mathbb{C}$@374};
\end{pgfonlayer}

\end{tikzpicture}}
	\end{minipage}
	}
	\caption{Computational efficiency in MFLOP/RD-map of RVCNNs and CVCNNs. F1-Score and EVM over MFLOP/RD-map were used as performance metrics. The best real-valued and complex-valued models are annotated with the used MFLOPS/RD-map.}
	\label{fig:rd-denoising_flops}
\end{figure}
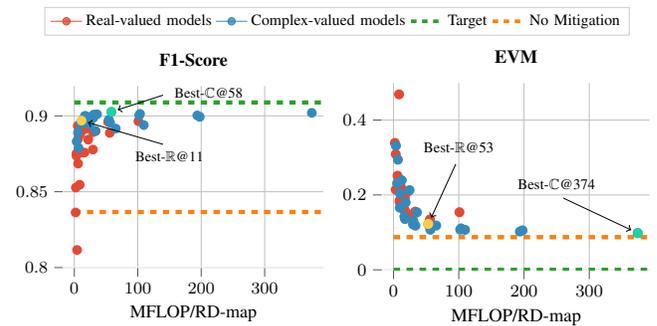
Comparing the results of the complex-valued network in Fig.~\ref{fig:rd-denoising_flops} with Fig.~\ref{fig:rd-denoising_param}, we see that the improvement in performance compared to the real-valued networks is not as pronounced as in the last experiment. This means that while being very parameter efficient, the complex-valued models introduce a lot of computational overhead. Therefore, the performance improvement on a per FLOP level is smaller than for the case of only considering parameter count. Nevertheless, the complex-valued models are able to outperform their real-valued counterparts for most architectures.

\subsection{Data efficiency}\label{sec:data_efficiency}
In this experiments we vary the used training data with a step size of one percent. The total number of training samples (=RD-maps) is 2500, which gives a minimum number and step size of 25 training samples.
We compare a three layer real-valued network containing 16-8-2 channels ($\mathbb{R}$ 16-8-2) with two variants of complex-valued networks ($\mathbb{C}$ 8-8-1 and $\mathbb{C}$ 8-4-1). The two complex-valued network variants are selected, such that they have a similar number of real-valued parameters (i.e. $\mathbb{C}$ 8-8-1) and a similar computational complexity (i.e. $\mathbb{C}$ 8-4-1) when compared to the real-valued baseline. See Section~\ref{sec:computational_complexity} and in particular Fig.~\ref{fig:flops_vs_params} for details about the parameter-to-FLOP/RD-map relation of real-valued and complex-valued networks. The figure indicates the closest complex-valued architectures with respect to both considered complexity measures, i.e. parameters and FLOP/RD-map.
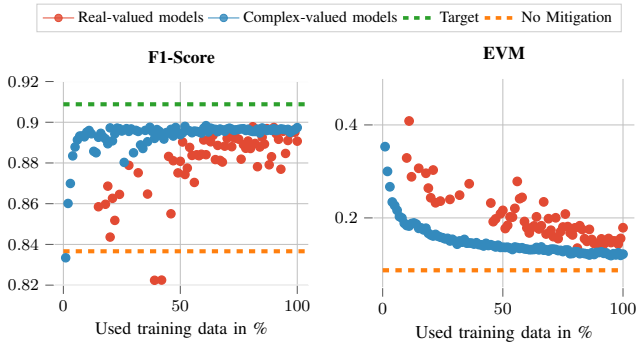
\begin{figure}[h]
{
\small
	\begin{minipage}[h]{0.9\columnwidth}
	    \hspace{0.05\textwidth}
		\resizebox {\textwidth} {!} {
            \begin{tikzpicture}

\definecolor{color0}{rgb}{0.886274509803922,0.290196078431373,0.2}
\definecolor{color1}{rgb}{0.203921568627451,0.541176470588235,0.741176470588235}
\definecolor{color2}{rgb}{0.976,0.859,0.839}
\definecolor{interferedorange}{rgb}{1,0.498039215686275,0.0549019607843137}
\definecolor{cleangreen}{rgb}{0.172549019607843,0.627450980392157,0.172549019607843}

\centering
\begin{axis}[%
hide axis,
xmin=0,
xmax=1,
ymin=0,
ymax=1,
scale=0.5,
legend columns=4,
legend style={fill opacity=0.8, draw opacity=1, text opacity=1, draw=white!80!black},
]
\addlegendimage{color0,mark=*}
\addlegendentry{Real-valued models}
\addlegendimage{color1,mark=*}
\addlegendentry{Complex-valued models}
\addlegendimage{cleangreen,dashed, line width=2pt}
\addlegendentry{Target}
\addlegendimage{interferedorange, dashed, line width=2pt}
\addlegendentry{No Mitigation}
\end{axis}
\end{tikzpicture}}
            \vspace{-1.8cm}
	\end{minipage}
	\begin{minipage}[h]{0.49\columnwidth}
		\resizebox {\textwidth} {!} {
\begin{tikzpicture}

\definecolor{color0}{rgb}{0.886274509803922,0.290196078431373,0.2}
\definecolor{color1}{rgb}{0.203921568627451,0.541176470588235,0.741176470588235}
\definecolor{color2}{rgb}{0.976,0.859,0.839}
\definecolor{interferedorange}{rgb}{1,0.498039215686275,0.0549019607843137}
\definecolor{cleangreen}{rgb}{0.172549019607843,0.627450980392157,0.172549019607843}

\begin{axis}[
scale only axis=true,
height=0.8\columnwidth,
width=1.0\columnwidth,
axis line style={white},
legend cell align={left},
legend style={fill opacity=0.8, draw opacity=1, text opacity=1, draw=white!80!black, fill=white!89.8039215686275!black},
tick align=outside,
tick pos=left,
title={\textbf{F1-Score}},
xlabel={Used training data in \%},
xmajorgrids,
xmin=-3.95, xmax=104.95,
xtick style={color=white!33.3333333333333!black},
ymajorgrids,
ymin=0.82, ymax=0.92,
ytick style={color=white!33.3333333333333!black}
]
\addplot [semithick, color0, mark=*, mark size=2, mark options={solid}, only marks]
table {%
1 0.395935982465744
2 0.367338001728058
3 0.277245551347733
4 0.582135319709778
6 0.31718048453331
7 0.339572042226791
10 0.767196595668793
11 0.736607313156128
12 0.808893024921417
13 0.74214768409729
14 0.486461341381073
15 0.858490824699402
18 0.859715819358826
19 0.868617177009583
20 0.843602001667023
21 0.862634420394897
22 0.851796984672546
23 0.778205871582031
24 0.864567279815674
27 0.765164434909821
28 0.87883198261261
32 0.875174403190613
35 0.787494301795959
36 0.864727973937988
38 0.802498400211334
39 0.822383999824524
42 0.82248443365097
44 0.700996935367584
45 0.883213698863983
46 0.855036854743958
47 0.881083786487579
49 0.875126302242279
50 0.880822002887726
51 0.890291690826416
52 0.874341487884521
53 0.877503156661987
54 0.88765823841095
55 0.883695483207703
56 0.870424151420593
57 0.88403582572937
58 0.883547365665436
59 0.884291052818298
60 0.891273260116577
61 0.884060919284821
62 0.891310632228851
63 0.890383839607239
64 0.893542468547821
65 0.881658911705017
66 0.888617336750031
67 0.881291270256042
68 0.880442142486572
69 0.88825535774231
70 0.892078876495361
71 0.894992411136627
72 0.887985587120056
73 0.892306089401245
74 0.892122864723206
75 0.891667246818542
76 0.881873786449432
77 0.888981997966766
78 0.887206196784973
79 0.890792012214661
80 0.887093424797058
81 0.897800862789154
82 0.889607965946198
83 0.878160834312439
84 0.887832701206207
85 0.892656624317169
86 0.889287948608398
87 0.896880388259888
88 0.879094541072845
89 0.883072972297668
90 0.897434234619141
91 0.894339978694916
92 0.891618013381958
93 0.876934170722961
94 0.895542979240417
95 0.884677052497864
96 0.891029357910156
97 0.894582152366638
98 0.895982444286346
99 0.895108819007874
100 0.890703916549683
};
\addplot [semithick, color1, mark=*, mark size=2, mark options={solid}, only marks]
table {%
1 0.833430588245392
2 0.860118746757507
3 0.86992347240448
4 0.883479237556458
5 0.887833595275879
6 0.891348898410797
7 0.89324426651001
8 0.893290638923645
9 0.892819166183472
10 0.895131051540375
11 0.895967602729797
12 0.894319713115692
13 0.885730922222137
14 0.884993195533752
15 0.89260458946228
16 0.894572913646698
17 0.892784714698792
18 0.891837239265442
19 0.889426469802856
20 0.897223830223083
21 0.890862941741943
22 0.894327759742737
23 0.897145390510559
24 0.896827220916748
25 0.895675301551819
26 0.880279123783112
27 0.896866083145142
28 0.8955078125
29 0.896719515323639
30 0.884989440441132
31 0.896022617816925
32 0.895473837852478
33 0.888873636722565
34 0.887109994888306
35 0.896000564098358
36 0.890430331230164
37 0.897540807723999
38 0.895483374595642
39 0.892056941986084
40 0.896421551704407
41 0.895584762096405
42 0.893337726593018
43 0.891729533672333
44 0.894926965236664
45 0.894636690616608
46 0.896517038345337
47 0.893210470676422
48 0.897458374500275
49 0.894849359989166
50 0.893877506256104
51 0.894785046577454
52 0.897983551025391
53 0.896116375923157
54 0.895775139331818
55 0.89483255147934
56 0.895683944225311
57 0.895522713661194
58 0.895215511322021
59 0.894677937030792
60 0.897091209888458
61 0.898336589336395
62 0.895233273506165
63 0.897122025489807
64 0.896614015102386
65 0.896517753601074
66 0.895383656024933
67 0.896687090396881
68 0.89628654718399
69 0.89555037021637
70 0.896663904190063
71 0.897111177444458
72 0.896612465381622
73 0.895976424217224
74 0.89632260799408
75 0.896531343460083
76 0.896716296672821
77 0.896818280220032
78 0.895956993103027
79 0.896592915058136
80 0.895770788192749
81 0.89544403553009
82 0.895438253879547
83 0.89665162563324
84 0.894715070724487
85 0.897191107273102
86 0.896602749824524
87 0.895953416824341
88 0.896404683589935
89 0.89599871635437
90 0.896886646747589
91 0.895866751670837
92 0.896279215812683
93 0.896736681461334
94 0.896568715572357
95 0.896301031112671
96 0.896518290042877
97 0.895009934902191
98 0.895875871181488
99 0.896411299705505
100 0.897376298904419
};

\addplot [dashed, line width=2pt, cleangreen]
table {%
0 0.908827448042987
10000 0.908827448042987
};
\addplot [dashed, line width=2pt, interferedorange]
table {%
0 0.836633657524875
10000 0.836633657524875
};
\end{axis}

\end{tikzpicture}}
	\end{minipage}%
	\begin{minipage}[h]{0.49\columnwidth}
		\resizebox {\textwidth} {!} {
\begin{tikzpicture}

\definecolor{color0}{rgb}{0.886274509803922,0.290196078431373,0.2}
\definecolor{color1}{rgb}{0.203921568627451,0.541176470588235,0.741176470588235}
\definecolor{color2}{rgb}{0.976,0.859,0.839}
\definecolor{interferedorange}{rgb}{1,0.498039215686275,0.0549019607843137}
\definecolor{cleangreen}{rgb}{0.172549019607843,0.627450980392157,0.172549019607843}

\begin{axis}[
scale only axis=true,
height=0.8\columnwidth,
width=1.0\columnwidth,
axis line style={white},
legend cell align={left},
legend style={fill opacity=0.8, draw opacity=1, text opacity=1, draw=white!80!black, fill=white!89.8039215686275!black},
tick align=outside,
tick pos=left,
title={\textbf{EVM}},
xlabel={Used training data in \%},
xmajorgrids,
xmin=-3.95, xmax=104.95,
xtick style={color=white!33.3333333333333!black},
ymajorgrids,
ymin=0.05, ymax=0.5,
ytick style={color=white!33.3333333333333!black}
]
\addplot [semithick, color0, mark=*, mark size=2, mark options={solid}, only marks]
table {%
1 0.995359420776367
2 0.989209055900574
3 0.972919583320618
4 0.61536180973053
5 1.00061440467834
6 0.837150752544403
7 0.833103895187378
8 0.995798707008362
9 0.985134959220886
10 0.328966468572617
11 0.408455371856689
12 0.288447976112366
13 0.682496726512909
14 0.800154805183411
15 0.306715786457062
18 0.296048253774643
19 0.264050930738449
20 0.243266761302948
21 0.302956521511078
22 0.232524618506432
23 0.689880132675171
24 0.236150816082954
27 0.69222366809845
28 0.239890903234482
32 0.248976111412048
35 0.684738636016846
36 0.273426234722137
38 0.683754205703735
39 0.644046545028687
42 0.642958879470825
44 0.708938956260681
45 0.23203906416893
46 0.191909432411194
47 0.197873741388321
49 0.208939909934998
50 0.216145515441895
51 0.177187860012054
52 0.184102445840836
53 0.202224984765053
54 0.201645180583
55 0.220454692840576
56 0.278494507074356
57 0.24135859310627
58 0.244107753038406
59 0.192256167531013
60 0.178601056337357
61 0.167707860469818
62 0.179049223661423
63 0.18277981877327
64 0.202353551983833
65 0.176300644874573
66 0.183143854141235
67 0.234462767839432
68 0.165384620428085
69 0.198356360197067
70 0.173504695296288
71 0.159716874361038
72 0.200284361839294
73 0.155448973178864
74 0.17039267718792
75 0.180401161313057
76 0.208343014121056
77 0.1812903881073
78 0.16776368021965
79 0.163371473550797
80 0.178586781024933
81 0.134963572025299
82 0.183162942528725
83 0.175239503383636
84 0.155269145965576
85 0.149629861116409
86 0.175320982933044
87 0.153548911213875
88 0.144982635974884
89 0.145212411880493
90 0.138028025627136
91 0.154868796467781
92 0.143552675843239
93 0.150952070951462
94 0.155696153640747
95 0.147006839513779
96 0.150538578629494
97 0.150120317935944
98 0.143770322203636
99 0.155879884958267
100 0.178889483213425
};

\addplot [semithick, color1, mark=*, mark size=2, mark options={solid}, only marks]
table {%
1 0.353253692388535
2 0.300072968006134
3 0.266952335834503
4 0.234156876802444
5 0.226565212011337
6 0.216367930173874
7 0.202875629067421
8 0.199823543429375
9 0.188492864370346
10 0.18375426530838
11 0.182542458176613
12 0.187807306647301
13 0.189106419682503
14 0.18663501739502
15 0.177614465355873
16 0.17845144867897
17 0.174344509840012
18 0.177835240960121
19 0.167019888758659
20 0.163393676280975
21 0.16114005446434
22 0.164444282650948
23 0.160692781209946
24 0.158619865775108
25 0.157227784395218
26 0.154980421066284
27 0.15597702562809
28 0.150459885597229
29 0.153929740190506
30 0.150176927447319
31 0.143507465720177
32 0.150410622358322
33 0.147873073816299
34 0.153660893440247
35 0.145784988999367
36 0.146105721592903
37 0.145756453275681
38 0.143797367811203
39 0.146507382392883
40 0.142703279852867
41 0.142035648226738
42 0.139698833227158
43 0.145215392112732
44 0.1407380849123
45 0.142278894782066
46 0.140206858515739
47 0.13762167096138
48 0.141199946403503
49 0.136615306138992
50 0.136825948953629
51 0.13650481402874
52 0.136939167976379
53 0.135034516453743
54 0.132082611322403
55 0.137306988239288
56 0.136185556650162
57 0.13449415564537
58 0.135770618915558
59 0.133477225899696
60 0.132804065942764
61 0.131726279854774
62 0.131823524832726
63 0.13220289349556
64 0.136419102549553
65 0.131358459591866
66 0.137453719973564
67 0.130661889910698
68 0.131490752100945
69 0.128808706998825
70 0.132696077227592
71 0.132865935564041
72 0.132054358720779
73 0.129818260669708
74 0.130152150988579
75 0.127653256058693
76 0.130782052874565
77 0.130855679512024
78 0.123027913272381
79 0.123598605394363
80 0.1274134516716
81 0.127056330442429
82 0.123104207217693
83 0.124156326055527
84 0.123453557491302
85 0.122548632323742
86 0.120267152786255
87 0.124793216586113
88 0.121389918029308
89 0.12571094930172
90 0.125697493553162
91 0.126954585313797
92 0.123868077993393
93 0.124789074063301
94 0.121069274842739
95 0.118958950042725
96 0.12288574129343
97 0.120486289262772
98 0.124971464276314
99 0.119523115456104
100 0.122136309742928
};


\addplot [dashed, line width=2pt, cleangreen]
table {%
0 0
100 0
};
\addplot [dashed, line width=2pt, interferedorange]
table {%
0 0.08736349517682467
100 0.08736349517682467
};

\end{axis}

\end{tikzpicture}}
	\end{minipage}
	}
	\caption{Data efficiency of a real-valued ($\mathbb{R}$ 16-8-2) and a complex-valued network ($\mathbb{C}$ 8-4-1) \emph{with similar computational complexity}. The F1-Score and EVM have been evaluated using different percentages (1\% to 100\%) of the training data.}
	\label{fig:data_8_4_1}
\end{figure}

TABLE~\ref{tab:data_efficiency} contains the F1-Score, EVM and PPMSE results for the two chosen network architectures ($\mathbb{C}$ 8-4-1 and $\mathbb{C}$ 8-8-1 complex channels) and the real-valued baseline model ($\mathbb{R}$ 16-8-2) for different fractions of used training data. Both complex-valued network architectures clearly outperform the real-valued equivalent, while the bigger complex-valued model $\mathbb{C}$ 8-8-1 typically yields the best performance scores. Fig.~\ref{fig:data_8_4_1} illustrates the data efficiency analysis of the smaller CVCNN architecture ($\mathbb{C}$ 8-4-1) and the real-valued baseline.

In general, the more training data we use, the better is the test performance for all three metrics (F1-Score, EVM and PPMSE). However, complex-valued networks require \emph{much} less training samples in order to yield high performances. While a complex-valued network reaches almost top results with only 10\% of the training samples, a comparable real-valued network requires at least 50\% of the training samples in order to train robustly and reach high performances. The performance difference is particularly high for metrics that consider phase information (EVM and PPMSE). This emphasizes the need for an accurate treatment of complex input domains in order to correctly infer complex information.

\begin{table}[h]
	\centering
	\begin{tabular}{c||c|c|c|c|c|c}
	    \textbf{Metric}&\textbf{Model}&\multicolumn{5}{c}{\textbf{Used data}}\\
	    \hline
		  & & 20~\% & 40~\% & 60~\%&80~\%&100~\% \\
		\hline
		\hline
		 \multirow{3}{*}{\textbf{F1-Score}}&$\mathbb{C}$ 8-8-1 & 0.895&\textbf{0.898} &0.896 &\textbf{0.898} & 0.897\\
        \cline{2-7}
        &$\mathbb{C}$ 8-4-1 &\textbf{0.897} & 0.896 & \textbf{0.897}&  0.896 & \textbf{0.897}\\
        \cline{2-7}
        &$\mathbb{R}$ 16-8-2 &0.872 &0.881 &0.887&0.891&0.885\\
        \hline
        \hline
		\multirow{3}{*}{\textbf{EVM}}&$\mathbb{C}$ 8-8-1 &\textbf{0.163} & \textbf{0.136} & \textbf{0.124} & \textbf{0.117} & \textbf{0.115}\\
		\cline{2-7}
		&$\mathbb{C}$ 8-4-1 &\textbf{0.163} &0.143&0.133 &0.127&0.122\\
		\cline{2-7}
		&$\mathbb{R}$ 16-8-2 &0.258&0.262 &0.209&0.184&0.146\\
		\hline
		\hline
		\multirow{3}{1 cm}{\textbf{PPMSE}\newline/~$\mathbf{rad^2}$}&$\mathbb{C}$ 8-8-1 & \textbf{0.017} & \textbf{0.014} & \textbf{0.013} & \textbf{0.013} & \textbf{0.013}\\
		\cline{2-7}
		&$\mathbb{C}$ 8-4-1 &0.018 &0.016&0.015 &0.014&0.014\\
		\cline{2-7}
		&$\mathbb{R}$ 16-8-2 &0.049&0.054 &0.035&0.022&0.018\\
		
	\end{tabular}
	\caption{\emph{F1-Score}, \emph{EVM} and \emph{PPMSE} results for different fractions of used training data. The RVCNN ($\mathbb{R}$ 16-8-2 channels) is compared to two CVCNNs ($\mathbb{C}$ 8-8-1 and $\mathbb{C}$ 8-4-1). The F1-Score without mitigation and from the target measurements is $\mathrm{F_{No Mit.}} = 0.837$ and $\mathrm{F_{Target}}=0.909$, respectively.}
	\label{tab:data_efficiency}
\end{table}

\subsection{Comparison with classical signal processing methods}\label{sec:comparison_classical}

\begin{figure*}[h]
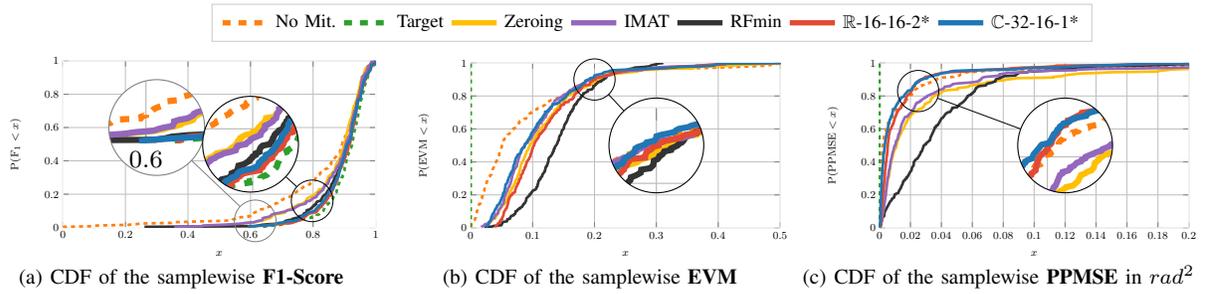

	\centering
	\footnotesize
	\begin{minipage}[h]{1.0\textwidth}
	    \hspace{0.1\textwidth}
	    \centering
            \begin{tikzpicture}

\definecolor{mycolor1}{rgb}{0.00000,0.00000,0.17241}%
\definecolor{mycolor2}{rgb}{1.00000,0.10345,0.72414}%

\definecolor{color2}{rgb}{1, 0.749, 0} 
\definecolor{color0}{rgb}{0.2,0.2,0.2} 
%

\definecolor{imatlila}{rgb}{0.580392156862745,0.403921568627451,0.741176470588235}
\definecolor{rvcnnred}{rgb}{0.886274509803922,0.290196078431373,0.2}
\definecolor{cvcnnblue}{rgb}{0.12156862745098,0.466666666666667,0.705882352941177}
\definecolor{cleangreen}{rgb}{0.172549019607843,0.627450980392157,0.172549019607843}
\definecolor{interferedorange}{rgb}{1,0.498039215686275,0.0549019607843137}
%
%
%

\centering
\begin{axis}[%
hide axis,
xmin=0,
xmax=1,
ymin=0,
ymax=1,
scale=0.5,
legend columns=7,
legend style={fill opacity=0.8, draw opacity=1, text opacity=1, draw=white!80!black, font=\scriptsize},
]
\addlegendimage{dashed,color=interferedorange, line width=2.0pt}
\addlegendentry{No Mit.}
\addlegendimage{dashed,color=cleangreen, line width=2.0pt}
\addlegendentry{Target}
\addlegendimage{color=color2, line width=2.0pt}
\addlegendentry{Zeroing}
\addlegendimage{color=imatlila, line width=2.0pt}
\addlegendentry{IMAT}
\addlegendimage{color=color0, line width=2.0pt}
\addlegendentry{RFmin}
\addlegendimage{semithick, rvcnnred,line width=2.0pt}
\addlegendentry{$\mathbb{R}$-${\text{16-16-2}}$*}
\addlegendimage{semithick, color=cvcnnblue,line width=2.0pt}
\addlegendentry{$\mathbb{C}$-${\text{32-16-1}}$*}


\end{axis}
\end{tikzpicture}
        \vspace{-2.3cm}
	\end{minipage}
	\subfigure[CDF of the samplewise \textbf{F1-Score}]{
		\resizebox {0.3\textwidth} {!} {
                \input{F1.tex}
	    }
		\label{fig:cdf_classical_f1}
	}
	\hspace{-0.4cm}
	\subfigure[CDF of the samplewise \textbf{EVM}]{
		\resizebox {0.3\textwidth} {!} {
                \input{EVM.tex}
	    }
		\label{fig:cdf_classical_evm}
	}
	\hspace{-0.4cm}
	\subfigure[CDF of the samplewise \textbf{PPMSE} in $rad^2$]{
		\resizebox {0.3\textwidth} {!} {
                \input{PPMSE.tex}
	    }
	\label{fig:cdf_classical_ppmse}
	}
	\caption{Empirical CDF performance comparison between the real-valued model\textsuperscript{*} ($\mathbb{R}$-model), the complex-valued model\textsuperscript{*} ($\mathbb{C}$-model), and the three classical methods zeroing, IMAT and Ramp filtering. *Model selected based on the best EVM result in the design space.}
	\label{fig:cdf_classical}
\end{figure*}

We compare our best CNN-based models in the real-valued setting ($\mathbb{R}$-model) as well as in the complex-valued setting ($\mathbb{C}$-model) with the classical and state-of-the-art interference mitigation methods zeroing \cite{Fischer}, \emph{Iterative method with adaptive thresholding (IMAT)} \cite{Bechter2017a} and \emph{Ramp filtering} \cite{WAG18}; see \cite{fu2017complex} for an overview of these methods. Zeroing and IMAT highly depend on an interference detection step, which influences their performance considerably. In our experiments we identified time-domain samples incorporating interference with approximately 90\% accuracy; this interference detection rate seems feasible in practice. Note that Ramp filtering as well as the CNN-based models do not depend on such an explicit interference detection step.
The average results for all methods are given in TABLE \ref{tab:param_efficiency}.  
Fig.~\ref{fig:cdf_classical} shows the empirical \emph{cumulative density function (CDF)} using the per-sample F1-Score and the EVM at peak locations. The 'clean' measurement and interfered signals are included as reference.
All classical methods, namely zeroing, IMAT and Ramp filtering, improve the F1-Score (as shown in Fig.~\ref{fig:cdf_classical_f1}) when applied to the measurement signal with interference. Zeroing and IMAT yield very similar F1-Scores. Ramp filtering outperforms both other classical methods, particularly for samples with strong interference (see black magnification).

The CNN-based models, both with real-valued as well as complex-valued weights and activations, are competitive with the classical methods for all considered interference levels and even outperform the best classical method, namely Ramp filtering (see black magnification). The CDF shape indicates that the CNN-based models are robust with respect to different interference patterns and levels. This is indicated by the CDF's narrow form and high values for the lowest F1-Scores per CNN model (see gray magnification). Note, that the F1-Score, and thus the detection sensitivity, of both CNN-based models is very close to the theoretical maximum, namely the CDF of the clean data.

Fig.~\ref{fig:cdf_classical_evm} shows the empirical CDF of the EVM. Generally, the better the mitigation in terms of F1-Score, the higher is the EVM and thus the distortion of the phase at object peaks. In comparison to the RVCNN, the CVCNN reduces the peak distortions although it achieves a very similar F1-Score and it even outperforms the best classical method in terms of EVM, namely IMAT. Hence, the CVCNN achieves very high F1-Scores while also retaining low EVMs. The PPMSE in Fig.~\ref{fig:cdf_classical_ppmse} shows similar characteristics. The CVCNN outperforms all other methods incorporating less phase distortions, particularly for weak interferences.

\begin{table}[h]
    \centering
    \scriptsize
    \begin{tabular}{c|c|c|c|c|r}
         & \textbf{F1-Score}& \textbf{EVM} & \textbf{PPMSE}&\textbf{Params} & \textbf{MFLOP}\\
         \hline
        $\mathbb{R}$-${\text{16-16-2}}$* &0.896&0.123 &0.015&2946&53.24\\
        \hline
		$\mathbb{C}$-${\text{32-16-1}}$* &\textbf{0.902} &\textbf{0.098} &\textbf{0.011}&10242&374.27\\
		\hline
		Zeroing & 0.856  & 0.124  & 0.069  & - & \textbf{2.83} \\
		\hline
		RFmin & 0.875  & 0.145  & 0.054  & - & 3.13 \\
		\hline
		IMAT & 0.860  & 0.119  & 0.065  & - & 3.88 \\
    \end{tabular}
    \caption{Performance comparison of real- and complex-valued models\textsuperscript{*} with the three classical methods zeroing, IMAT and Ramp filtering in terms of F1-Score, EVM and PPMSE. The F1-Score without mitigation and from the target measurements is $\mathrm{F_{No Mit.}}=0.837$ and $\mathrm{F_{Target}}=0.909$, respectively. *Model selected based on the best EVM result in the design space.
    }
    \label{tab:param_efficiency}
\end{table}

\section{Conclusion}\label{sec:conclusion}
We propose the use of complex-valued CNNs for range-Doppler denoising and interference mitigation of automotive radar signals. The proposed NN architecture follows complex-valued analysis and thus processes the complex-valued signals according to their physical characteristics. This inductive bias restricts the network structure and operations in a meaningful way and therefore reduces the complexity of the learning problem. We confirm this claim with experiments on complex-valued radar signals. We show, that complex-valued networks with a similar (1) number of parameters and (2) computational complexity substantially improve all considered metrics (F1-Score, EVM and PPMSE) and that they are able to operate on \emph{much} smaller training sets. Particularly the high performance of metrics considering the phase information (EVM and PPMSE) reveals the advantage of complex-valued NNs for applications on complex-valued signal domains. The proposed approach might thus be crucial for processing complex-valued signals, such as automotive radar signals containing valuable phase information, e.g. for angle estimation. In future research we want to focus on the transferability of models between different radar antennas, the impact of multi-antenna information during training and the direct evaluation of angle estimation capabilities.




%
\bibliographystyle{ieeetr}
\bibliography{refs}

\end{document}